\documentstyle[amssymb,12pt]{article}

\def\pur#1(#2){\nu_{#1}(#2)}
\def\puri(#1){\pur\infty(#1)}
\def\purS(#1){\pur H(#1)}

\begin{document}

\title{On the multiplicativity conjecture for quantum channels}
\author{G. G. Amosov, A. S. Holevo}
\abstract{A multiplicativity conjecture for quantum communication channels is
formulated, validity of which for the values of parameter $p$ close to 1
is related to the solution of the fundamental problem of additivity of
the channel capacity in quantum information theory. The proof of the
conjecture is given for the case of natural numbers $p$.}

\maketitle

1. Let ${\cal B(H)}$ be the $*$ -algebra of all operators in a finite
dimensional unitary space ${\cal H}$. We denote the set of {\it states},
i.e. positive operators with unit trace (density operators) in ${\cal B(H)}$
by ${\cal S(H)}$.
As is well known,
${\cal S(H)}$ is a compact convex subset of
${\cal B(H)}$, extreme points of which are {\it pure states},
described by one-dimensional projectors in
 ${\cal H}$.  The degree of ``purity'' of arbitrary state $S\in{\cal S(H)}$
can be defined with the help of noncommutative $\ell _{p}$-norms \[ \Vert
S\Vert _{p}=\left( {\rm Tr}|S|^{p}\right) ^{\frac{1}{p}},\quad p\geq 1, \]
with the operator norm $\Vert
S\Vert $ corresponding naturally to the case $p=\infty $.  The closer is the
value of any norm to the identity, the more ``pure'' is the state $S$.

A {\it quantum channel} $\Phi $ is a completely positive
trace preserving linear map of ${\cal B(H)}$, i.e. a map admitting the
representation
\begin{equation}
\Phi (S )=\sum_{k}A_{k}S A_{k}^{*},  \label{kra}
\end{equation}
where $A_{k}$ are operators satisfying $\sum_{k}A_{k}^{*}A_{k}=I$ (see e. g.
\cite{hol} for motivation and background). The channel $\Phi $
maps input state $S$ into output state $\Phi (S)$. In the present paper we
consider the multiplicativity problem for the measures of the
``highest purity'' of outputs of a channel
\begin{equation}
\pur p(\Phi )=\max_{S }\Vert \Phi (S )\Vert _{p},  \label{mi}
\end{equation}
where the maximum is taken with respect to all input density operators $S$.
By convexity of the norms, the maximum in the above definition is
attained on pure states.

Let $\Phi _{1},\dots ,\Phi _{n}$ be a collection of arbitrary channels in
the unitary spaces ${\cal H}_{i};\ i=1,2,...,n$. In \cite{ahw} the following
multiplicativity property
\begin{equation}
\pur p(\Phi _{1}\otimes \dots \otimes \Phi _{n})=\pur p(\Phi
_{1})\cdot\dots\cdot \pur p(\Phi _{n})  \label{ae} \end{equation} was
conjectured.  As noticed in \cite{ahw}, validity of this conjecture for
values of $p$  close to 1 implies solution of the fundamental  problem of
{\it additivity of the channel capacity} for one important class of quantum
channels. For the formulation of the and some partial results see
\cite{BS}, \cite{ahw}, \cite{rus}.  In the classical case where ${\cal B(H)}$
is replaced by a commutative algebra of diagonal operators, the states are
given by probability distributions, and channels -- by transition
probabilities, the analog of the formulated additivity/multiplicativity
problems has obvious positive solution. The difficulty in the
noncommutative case is due to the unusual from a classical viewpoint
properties of combined quantum systems described by tensor rather
than Cartesian products, and by existence of entangled states in the combined
system.

2. We denote by $\ell
_{p}({\cal H})$ the Schatten class of {\it Hermitian} operators
$A$ in $ {\cal H}$ with the norm $\Vert A\Vert _{p}.$

{\bf Lemma. }{\it The quantity }$\pur p(\Phi )$ {\it is equal to the norm
}$ \Vert \Phi \Vert _{1\rightarrow p}$ {\it of the mapping }$\Phi ${\it \
acting from the Schatten class }$\ell _{1}({\cal H})$ {\it to }$\ell _{p}(
{\cal H}).$

{\it Proof. }We have
\[
\Vert \Phi \Vert _{1\rightarrow p}=\max_{A\neq 0}\frac{\left( {\rm Tr}|\Phi
(A)|^{p}\right) ^{\frac{1}{p}}}{{\rm Tr}|A|},
\]
so obviously $\Vert \Phi \Vert _{1\rightarrow p}\ge \pur p(\Phi ).$
Conversely, let $A=A_{+}-A_{-}$ be the decomposition of $A$ into positive
and negative parts, then ${\rm Tr}|A|={\rm Tr}${\rm (}$A_{+}+A_{-})$ and $%
-\Phi (A_{+}+A\_)\le \Phi (A)\le \Phi (A_{+}+A\_)$ by positivity of $\Phi .$
From convexity of the function $x^{p},{\rm Tr}|\Phi (A)|^{p}\le {\rm Tr}\Phi
(A_{+}+A_{-})^{p},$  indeed, denoting $\left\{ e_{j}\right\} $ the basis of
eigenvectors of $\Phi (A),$ we have
\[
{\rm Tr}|\Phi (A)|^{p}=\sum_{j}\left| \langle e_{j}|\Phi (A)|e_{j}\rangle
\right| ^{p}\le \sum_{j}\langle e_{j}|\Phi (|A|)|e_{j}\rangle ^{p}\]\[\le
\sum_{j}\langle e_{j}|\Phi (|A|)^{p}|e_{j}\rangle ={\rm Tr}\Phi (|A|)^{p},
\]
and the converse inequality follows.$\Box $

Generalizing hypothesis (\ref{ae}), we conjecture that the norms $\Vert \Phi
\Vert _{q\rightarrow p}$ have a similar multiplicative property for $1\le
q\le p$ for  completely positive maps $\Phi _{1},\dots ,\Phi _{n}.$ Note
that the classical (commutative) counterpart of this conjecture indeed holds
for arbitrary (bounded) $\Phi _{1},\dots ,\Phi _{n}$ (\cite{beck}, Lemma 2)$.
$

In \cite{ahw} relation (\ref{ae}) was proved for the special case of {\it %
depolarizing channels} $\Phi _{1},\dots ,\Phi _{n}$ (see the definition in
n.3) and $p=2,\infty .$ Here this property will be established for the
depolarizing channels and arbitrary natural number $p.$

3. Let us consider a collection of unitary spaces ${\cal H}_{i};\
i=1,2,...,n,$ with ${\rm dim}{\cal H}_{i}=d_{i}.$ Let ${\cal H}=\otimes
_{i=1}^{n}{\cal H}_{i}$, $d=\prod_{i=1}^{n}d_{i}.$ In what follows we shall
use symbols $I_{i}$ and $I=\otimes _{i=1}^{n}I_{i}$ for the identity
operators in ${\cal H}_{i}$ and ${\cal H}$, respectively. For a nonempty
subset $L\subset \{1,2,...,n\}$ we denote
\[
{\cal H}_{L}=\otimes _{i\in L}{\cal H}_{i},\quad I_{L}=\otimes _{i\in
L}I_{i},\quad d_{L}=\prod_{i\in L}d_{i}={\rm dim}{\cal H}_{L}.
\]
We also let $d_{\emptyset }=1.$

Let $L_{1},\dots , L_{m}$ be a collection of nonempty subsets of $
\{1,2,...,n\},$ and let $A_{1},\dots ,A_{m}$ be a collection of operators in $
{\cal H}$ such that $A_{k}=B_{k}\otimes I_{L_{k}},$ where $B_{k}$ is an
operator in ${\cal H}_{L_{k}^{c}}.$

 {\bf Lemma.}
\begin{equation}
|{\rm Tr}A_{1}\dots A_{m}|\leq d_{\bigcap_{k}L_{_{k}}}||B_{1}||_{1}\dots
||B_{m}||_{1}.  \label{lemma}
\end{equation}

{\it Proof.} By using the singular value decomposition of the operators $%
B_{k},$ we can reduce the problem to the case where $B_{k}$ are the rank one
operators, $B_{k}=|a_{k}\rangle \langle b_{k}|$ with unit vectors $%
|a_{k}\rangle ,|b_{k}\rangle .$ Moreover, by excluding the common factor $%
I_{_{\bigcap_{k}L_{_{k}}}},$ we can reduce to the case $%
\bigcap_{k=1}^{m}L_{k}=$ $\emptyset .$ Then (\ref{lemma}) reduces to{\it
\begin{equation}
|{\rm Tr}A_{1}\dots A_{m}|\leq 1.  \label{remma}
\end{equation}
} Pick an orthonormal basis $\left\{ e_{j_{s}}\right\} $ in ${\cal H}_{s},$
and form the factorizable basis $\left\{ e_{J}\right\} $ in the space ${\cal %
H}$ such that
\[
e_{J}=e_{j_{1}}\otimes e_{j_{2}}\otimes \dots \otimes e_{j_{n}},
\]
where $J=(j_{1},j_{2},\dots ,j_{n})$. Denote $J_{L}=(j_{s})_{s\in L}.$ Then
by decomposing the unit vectors $|a_{k}\rangle ,|b_{k}\rangle ,$ we have
\[
|a_{k}\rangle =\sum_{J_{L_{k}^{c}}}\alpha _{J_{L_{k}^{c}}}^{k}|e_{J_{\bar{L}%
_{k}}}\rangle ,\quad |b_{k}\rangle =\sum_{J_{L_{k}^{c}}}\beta
_{J_{L_{k}^{c}}}^{k}|e_{J_{\bar{L}_{k}}}\rangle ,
\]
where $\sum_{J_{L_{k}^{c}}}|\alpha
_{J_{L_{k}^{c}}}^{k}|^{2}=\sum_{J_{L_{k}^{c}}}|\beta
_{J_{L_{k}^{c}}}^{k}|^{2}=1,$ so that
\begin{equation}
A_{k}=\sum_{J_{L_{k}^{c}}}\alpha
_{J_{L_{k}^{c}}}^{k}|e_{J_{L_{k}^{c}}}\rangle \sum_{J_{L_{k}^{c}}^{\prime
}}\beta _{J_{L_{k}^{c}}^{\prime }}^{k}\langle e_{J_{L_{k}^{c}}^{\prime
}}|\otimes \sum_{J_{L_{k}}}|e_{J_{L_{k}}}\rangle \langle e_{J_{L_{k}}}|.
\label{xk}
\end{equation}

Now consider the multiindex ${\frak J}=\left( J_{L_{k}^{c}}\right)
_{k=1,\dots ,m},$ the components of which are labeled by pairs $\left(
k,p\right) ,$ where $k\in \left\{ 1,\dots m\right\} $ and $p\in L_{k}^{c},$
and put
\[
\alpha _{{\frak J}}=\prod_{k=1}^{m}\alpha _{J_{L_{k}^{c}}}^{k},\quad \beta _{%
{\frak J}}=\prod_{k=1}^{m}\beta _{J_{L_{k}^{c}}}^{k}.
\]
Then
\begin{equation}
\sum_{{\frak J}}|\alpha _{{\frak J}}|^{2}=\sum_{{\frak J}}|\beta _{{\frak J}%
}|^{2}=1.  \label{norm}
\end{equation}
We are going to show that substituting (\ref{xk}) into the left side of (\ref
{remma}), we obtain
\begin{equation}
{\rm Tr}A_{1}\dots A_{m}=\sum_{{\frak J}}\bar{\beta}_{{\frak J}}\alpha _{%
{\cal P}{\frak J}},  \label{cs}
\end{equation}
where ${\cal P}{\frak J}$ is a permutation of the components of the
multiindex ${\frak J},$ and hence by the Cauchy-Schwarz inequality and (\ref
{norm}) we have the inequality (\ref{remma}).

The permutation ${\cal P}$ arises as follows: let us take $A_{k}$ of the
form (\ref{xk}) in the expression ${\rm Tr}$ $A_{1}\dots A_{m},$ and let $%
s\in L_{k}^{c}.$ Then the covectors $\langle e_{j_{s}}|$ are present in the
decomposition of $\langle b_{k}|.$ Let us go right cyclically under the
trace starting from $A_{k},$ passing through the identity operators, and
watch when $s$ will first again appear in $\bar{L}_{k\boxplus l},$ where $%
\boxplus $ denotes addition ${\rm {mod}m.}$ Then the vectors $%
|e_{j_{s}}\rangle $ will be present in the decomposition of $|a_{k\boxplus
l}\rangle ,$ giving rise to the summation over $j_{s}$ in (\ref{cs}).

More precisely, let us denote by ${\cal A}$ the set of all the pairs $\left(
k,s\right) ,$ where $k\in \left\{ 1,\dots m\right\} $ and $s\in \bar{L}_{k}.$
Thus the components of ${\frak J}$ can be written as $j(k,s),$ where $\left(
k,s\right) \in {\cal A}$. Let $l$ be the minimal ${\rm {mod}}$ $m$ positive
integer such that $\left( k\boxplus l,s\right) \in {\cal A}$. The mapping $%
\left( k,s\right) \rightarrow \left( k\boxplus l,s\right) $ is a bijection
of the set ${\cal A}$, therefore it induces a permutation ${\cal P}$ of the
multiindex ${\frak J}$, resulting in the formula (\ref{cs}).$\Box $

4. Let us consider a collection of depolarizing channels
\begin{equation}
\Phi _{i}(S )=(1-p_{i})S +\frac{p_{i}}{d_{i}}(\mbox{Tr}S )I_{i},\
S \in {\cal B(H}_{i}{\cal )},\ 0<p_{i}<1,  \label{dep}
\end{equation}
in the unitary spaces ${\cal H}_{i},$ with parameters $p_{i},d_{i};\
i=1,2,...,n,$ and denote $\Phi =\otimes _{i=1}^{n}\Phi _{i}$. It is easy to
see that $\Phi_i$ are indeed channels, i.e. can be represented in the form
(\ref{kra}). If $S $ is
pure state (one-dimensional projection) then operator (\ref{dep}) has the
simple eigenvalue $(1-\frac{d_{i}-1}{d_{i}}p_{i})$ and $d_{i}-1$ eigenvalues
$\frac{p_{i}}{d_{i}}.$ Hence
\[
\pur{k}({\Phi }_{i})=\left[ \left( 1-(d_{i}-1)\frac{p_{i}}{d_{i}}\right)
^{p}+(d_{i}-1)\left( \frac{p_{i}}{d_{i}}\right) ^{p}\right] ^{\frac{1}{p}}.
\]

Let $\epsilon _{L}$ be the conditional expectation onto the subalgebra $
{\cal M}_{L^{c}}$, generated by operators of the form $A_{1}\otimes
...\otimes A_{n},$ where $A_{i}=I_{i}$ for $i\in L,$ and arbitrary
otherwise. It is normalized partial trace with respect to ${\cal H}_{L}$:
\begin{equation}
\epsilon _{L}(A)={\rm Tr}_{{\cal H}_{L}}A\otimes d_{L}^{-1}I_{L}.
\label{epsi}
\end{equation}
In the following we shall use the expansion
\begin{equation}
\Phi =\sum\limits_{L}\prod\limits_{i=1}^{n}p_{i}^{\theta
_{L}(i)}(1-p_{i})^{1-\theta _{L}(i)}\epsilon _{L},  \label{B}
\end{equation}
where $\theta _{L}(i)=1$ if $i\in L$ and $\theta _{L}(i)=0$ otherwise. Note
that $d_{L}=\prod\limits_{i=1}^{n}d_{i}^{\theta _{L}(i)}$.

{\bf Theorem. }{\it For }$p\in {\bf N}$

$\quad \pur{k}(\Phi )=\prod\limits_{i=1}^{n}\pur{k}({\Phi }%
_{i})=\prod\limits_{i=1}^{n}\left[ \left( 1-(d_{i}-1)\frac{p_{i}}{d_{i}}%
\right) ^{p}+(d_{i}-1)\left( \frac{p_{i}}{d_{i}}\right) ^{p}\right] ^{\frac{1%
}{p}}\ {\bf .}$

{\it Proof. } Let $S \in $ ${\cal S(H)}$, then by (\ref{epsi}) and by (%
\ref{lemma})
\[
|{\rm Tr}\epsilon _{L_{1}}(S )\dots \epsilon _{L_{p}}(S )|\leq \frac{%
d_{\cap _{i}L_{i}}}{\prod_{i}d_{L_{i}}}.
\]
The obtained inequality and the expansion (\ref{B}) imply
\[
\mbox {Tr}\Phi (S )^{p}={\rm Tr}(\sum\limits_{L}\prod%
\limits_{i=1}^{n}p_{i}^{\theta _{L}(i)}(1-p_{i})^{1-\theta _{L}(i)}\epsilon
_{L}(S ))^{p}=
\]
\[
\sum\limits_{L_{1},L_{2},\dots
,L_{p}}\prod_{i=1}^{n}p_{i}^{\sum\limits_{j=1}^{p}\theta
_{L_{j}}(i)}(1-p_{i})^{\sum\limits_{j=1}^{p}(1-\theta _{L_{j}}(i))}\mbox{Tr}%
\epsilon _{L_{1}}(S )\dots \epsilon _{L_{p}}(S )\leq
\]
\[
\sum\limits_{L_{1},L_{2},\dots ,L_{p}}\prod_{i=1}^{n}\left( \frac{p_{i}}{%
d_{i}}\right) ^{\sum\limits_{j=1}^{p}\theta
_{L_{j}}(i)}(1-p_{i})^{\sum\limits_{j=1}^{p}(1-\theta
_{L_{j}}(i))}d_{i}^{\theta _{\bigcap_{j=1}^{p}L_{j}}(i)}.
\]
The number $\theta _{\bigcap_{j=1}^{p}L_{j}}(i)$ is equal to $\min \left\{
\theta _{L_{1}}(i),\dots ,\theta _{L_{p}}(i)\right\} $ and is equal to 1 if
and only if all $\theta _{L_{j}}(i);j=1,\dots ,p,$ are equal to 1. By using
the formula
\[
\sum\limits_{L_{1},L_{2},\dots ,L_{p}}\prod_{i=1}^{n}f_{i}\left( \theta
_{L_{1}}(i),\dots ,\theta _{L_{p}}(i)\right)
=\prod_{i=1}^{n}\sum\limits_{\theta _{1},\theta _{2},\dots ,\theta
_{p}=0,1}f_{i}\left( \theta _{1},\dots ,\theta _{p}\right) ,
\]
we obtain that the last expression is equal to
\[
\prod\limits_{i=1}^{n}\left[ \left( 1-(d_{i}-1)\frac{p_{i}}{d_{i}}\right)
^{p}+(d_{i}-1)\left( \frac{p_{i}}{d_{i}}\right) ^{p}\right] .\quad \Box
\]


\begin{thebibliography}{9}
\bibitem{ahw}  G. G. Amosov, A. S. Holevo and R. F. Werner. On some
additivity problems in quantum information theory. Probl. Inform. Transm.,
{\bf 36}, no.4, 25-34, 2000. LANL e-print math-ph/0003002.

\bibitem{beck}  W. Beckner, Inequalities in Fourier analysis, Ann. Math.,
{\bf \ 102}, 159-182, 1975.

\bibitem{BS} C. H.  Bennett, P. W. Shor, Quantum information
theory, IEEE Trans. Inform. Theory,  {\bf 44}, 2724-2742, 1998.

\bibitem{hol}  A.S. Holevo, Quantum coding theorems, Russian Math. Surveys
{\bf 53:6}, 1295-1331, 1998. LANL e-print quant-ph/9808023.

\bibitem{rus}  C.  King,  M. B. Ruskai,  Minimal entropy of states
emerging from noisy quantum channels, LANL e-print quant-ph/9911079.
\end{thebibliography}
\end{document}